\documentclass[longauth]{aa}  
\usepackage{natbib}
\usepackage[acronym]{glossaries}
\usepackage{graphicx}
\usepackage{txfonts}
\usepackage{hyperref}
\hypersetup{
    colorlinks=true,
    linkcolor=blue,
    citecolor=blue,
    }
\usepackage{xcolor}
\usepackage{textcomp,gensymb}

\titlerunning{ICM and dynamical state study of ACT-CL J0240.0+0116 using NIKA2 data}
\begin{document} 

    \title{Exploiting the high-resolution NIKA2 data to study the intracluster medium and dynamical state of ACT-CL J0240.0+0116}

 \author{A. Paliwal \inst{\ref{Roma}} \thanks{\email{aishwarya.paliwal@uniroma1.it}}    
     \and  M. De Petris \inst{\ref{Roma}}
     \and  A. Ferragamo \inst{\ref{Roma}}
     \and  R.~Adam \inst{\ref{OCA}}
     \and  P.~Ade \inst{\ref{Cardiff}}
     \and  H.~Ajeddig \inst{\ref{CEA}}
     \and  P.~Andr\'e \inst{\ref{CEA}}
     \and  E.~Artis \inst{\ref{Garching}}
     \and  H.~Aussel \inst{\ref{CEA}}
     \and  I. Bartalucci \inst{\ref{inaf_milan}}
     \and  A.~Beelen \inst{\ref{LAM}}
     \and  A.~Beno\^it \inst{\ref{Neel}}
     \and  S.~Berta \inst{\ref{IRAMF}}
     \and  L.~Bing \inst{\ref{LAM}}
     \and  O.~Bourrion \inst{\ref{LPSC}}
     \and  M.~Calvo \inst{\ref{Neel}}
     \and  A.~Catalano \inst{\ref{LPSC}}
     \and  F. De Luca \inst{\ref{ur2},\ref{inaf_roma}}
     \and  F.-X.~D\'esert \inst{\ref{IPAG}}
     \and  S.~Doyle \inst{\ref{Cardiff}}
     \and  E.~F.~C.~Driessen \inst{\ref{IRAMF}}
     \and  G.~Ejlali \inst{\ref{Teheran}}
     \and  A.~Gomez \inst{\ref{CAB}} 
     \and  J.~Goupy \inst{\ref{Neel}}
     \and  C.~Hanser \inst{\ref{LPSC}}
     \and  S.~Katsioli \inst{\ref{AthenObs}, \ref{AthenUniv}}
     \and  F.~K\'eruzor\'e \inst{\ref{Argonne}}
     \and  C.~Kramer \inst{\ref{IRAMF}}
     \and  B.~Ladjelate \inst{\ref{IRAME}} 
     \and  G.~Lagache \inst{\ref{LAM}}
     \and  S.~Leclercq \inst{\ref{IRAMF}}
     \and  J.-F.~Lestrade \inst{\ref{LERMA}}
     \and  J.~F.~Mac\'ias-P\'erez \inst{\ref{LPSC}}
     \and  S.~C.~Madden \inst{\ref{CEA}}
     \and  A.~Maury \inst{\ref{CEA}}
     \and  P.~Mauskopf \inst{\ref{Cardiff},\ref{Arizona}}
     \and  F.~Mayet \inst{\ref{LPSC}}
     \and  J.-B. Melin \inst{\ref{jbm}}
     \and  A.~Monfardini \inst{\ref{Neel}}
     \and  A.~Moyer-Anin \inst{\ref{LPSC}}
     \and  M.~Mu\~noz-Echeverr\'ia \inst{\ref{IRAP}}
     \and  L.~Perotto \inst{\ref{LPSC}}
     \and  G.~Pisano \inst{\ref{Roma}}
     \and  E. Pointecouteau \inst{\ref{IRAP}}
     \and  N.~Ponthieu \inst{\ref{IPAG}}
     \and  G.W. Pratt \inst{\ref{CEA}}
     \and  V.~Rev\'eret \inst{\ref{CEA}}
     \and  A.~J.~Rigby \inst{\ref{Leeds}}
     \and  A.~Ritacco \inst{\ref{ENS}, \ref{INAF}}
     \and  C.~Romero \inst{\ref{Pennsylvanie}}
     \and  H.~Roussel \inst{\ref{IAP}}
     \and  F.~Ruppin \inst{\ref{IP2I}}
     \and  K.~Schuster \inst{\ref{IRAMF}}
     \and  A.~Sievers \inst{\ref{IRAME}}
     \and  C.~Tucker \inst{\ref{Cardiff}}
     \and  R. Wicker \inst{\ref{Roma}}
     \and  R.~Zylka \inst{\ref{IRAMF}}
     }
   \institute{
     Dipartimento di Fisica, Sapienza Universit\`a di Roma, Piazzale Aldo Moro 5, I-00185 Roma, Italy
     \label{Roma}
     \and
     Universit\'e C\^ote d'Azur, Observatoire de la C\^ote d'Azur, CNRS, Laboratoire Lagrange, France 
     \label{OCA}
     \and
     School of Physics and Astronomy, Cardiff University, Queen’s Buildings, The Parade, Cardiff, CF24 3AA, UK 
     \label{Cardiff}
     \and
     Universit\'e Paris-Saclay, Universit\'e Paris Cit\'e, CEA, CNRS, AIM, 91191, Gif-sur-Yvette, France
     \label{CEA}
     \and
     Max Planck Institute for Extraterrestrial Physics, Giessenbach-strasse 1, 85748 Garching, Germany
     \label{Garching}
     \and
     INAF/IASF-Milano, Via A. Corti 12, 20133 Milano, Italy
     \label{inaf_milan}
     \and	
     Aix Marseille Univ, CNRS, CNES, LAM (Laboratoire d'Astrophysique de Marseille), Marseille, France
     \label{LAM}
     \and
     Institut N\'eel, CNRS, Universit\'e Grenoble Alpes, France
     \label{Neel}
     \and
     Institut de RadioAstronomie Millim\'etrique (IRAM), Grenoble, France
     \label{IRAMF}
     \and 
     Univ. Grenoble Alpes, CNRS, Grenoble INP, LPSC-IN2P3, 53, avenue des Martyrs, 38000 Grenoble, France
     \label{LPSC}
     \and
     Dipartimento di Fisica, Università di Roma ‘Tor Vergata’, Via della Ricerca Scientifica 1, I-00133 Roma, Italy
     \label{ur2}
     \and 
     INFN, Sezione di Roma `Tor Vergata’, Via della Ricerca Scientifica, 1, 00133 Roma, Italy
     \label{inaf_roma}
     \and
     Univ. Grenoble Alpes, CNRS, IPAG, 38000 Grenoble, France 
     \label{IPAG}
     \and
     Institute for Research in Fundamental Sciences (IPM), School of Astronomy, Tehran, Iran
     \label{Teheran}
     \and
     Centro de Astrobiolog\'ia (CSIC-INTA), Torrej\'on de Ardoz, 28850 Madrid, Spain
     \label{CAB}
     \and
     National Observatory of Athens, Institute for Astronomy, Astrophysics, Space Applications and Remote Sensing, Ioannou Metaxa
     and Vasileos Pavlou GR-15236, Athens, Greece
     \label{AthenObs}
     \and
     Department of Astrophysics, Astronomy \& Mechanics, Faculty of Physics, University of Athens, Panepistimiopolis, GR-15784
     Zografos, Athens, Greece
     \label{AthenUniv}
     \and
     High Energy Physics Division, Argonne National Laboratory, 9700 South Cass Avenue, Lemont, IL 60439, USA
     \label{Argonne}
     \and  
     Instituto de Radioastronom\'ia Milim\'etrica (IRAM), Granada, Spain
     \label{IRAME}
     \and
     LERMA, Observatoire de Paris, PSL Research University, CNRS, Sorbonne Universit\'e, UPMC, 75014 Paris, France  
     \label{LERMA}
     \and
     School of Earth and Space Exploration and Department of Physics, Arizona State University, Tempe, AZ 85287, USA
     \label{Arizona}
     \and
     IRFU, CEA, Université Paris-Saclay, 91191 Gif-sur-Yvette, France
     \label{jbm}
     \and
     IRAP, CNRS, Université de Toulouse, CNES, UT3-UPS, Toulouse, France
     \label{IRAP}
     \and
     School of Physics and Astronomy, University of Leeds, Leeds LS2 9JT, UK
     \label{Leeds}
     \and
     Laboratoire de Physique de l’\'Ecole Normale Sup\'erieure, ENS, PSL Research University, CNRS, Sorbonne Universit\'e, Universit\'e de Paris, 75005 Paris, France 
     \label{ENS}
     \and
     INAF-Osservatorio Astronomico di Cagliari, Via della Scienza 5, 09047 Selargius, IT
     \label{INAF}
     \and    
     Department of Physics and Astronomy, University of Pennsylvania, 209 South 33rd Street, Philadelphia, PA, 19104, USA
     \label{Pennsylvanie}
     \and
     Institut d'Astrophysique de Paris, CNRS (UMR7095), 98 bis boulevard Arago, 75014 Paris, France
     \label{IAP}
     \and
     University of Lyon, UCB Lyon 1, CNRS/IN2P3, IP2I, 69622 Villeurbanne, France
     \label{IP2I}
     }

\date{Received .....; accepted .....}

\abstract{Having a detailed knowledge of the intracluster medium (ICM) to infer the exact cluster physics such as the cluster dynamical state is crucial for cluster-based cosmological studies. This knowledge limits the accuracy and precision of mass estimation, a key parameter for such studies. In this paper, we conduct an in-depth analysis of cluster ACT-CL J0240.0+0116 using a multi-wavelength approach, with a primary focus on high angular resolution Sunyaev-Zeldovich (SZ) thermal component observations obtained under the NIKA2 Sunyaev-Zeldovich Large Programme (LPSZ). We create composite images using NIKA2, X-ray, and optical galaxy number density maps. The results reveal distinct signs of disturbance within the cluster with the distributions of gas and member galaxies that do not overlap. We also find suggestions of an inflow of matter onto the cluster from the southwestern direction. Ultimately, we classify the cluster as disturbed, using morphological indicators derived from its SZ, X-ray, and optical image. The cluster SZ signal is also contaminated by a strong central point source. We adopt different approaches to handling this contaminant and find the estimates of our pressure and hydrostatic mass profiles robust to the point source mitigation model. The cluster hydrostatic mass is estimated at $4.25^{+0.50}_{-0.45\, } \times 10^{14} \,\mathrm{M}_{\odot}$ for the case where the point source was masked. These values are consistent with the mass estimated using only X-ray data and with those from previous SZ studies of the Atacama cosmology telescope (ACT) survey, with improved precision on the mass estimate. Our findings strongly suggest that ACT-CL J0240.0+0116 is a disturbed cluster system, and the detailed observations and derived values serve as a compelling case study for the capabilities of the LPSZ in mapping the cluster ICM with high precision.}


   \maketitle
%
\newacronym{toi}{TOI}{Time-Ordered Information}
\newacronym{cm}{CM}{Common Mode}
\newacronym{gnfw}{gNFW}{generalized Navarro-Frenk-White}
\newacronym{psf}{PSF}{point spread function}
\newacronym{cib}{CIB}{cosmic infrared background}
\newacronym{np}{NP}{non-parametric}
\newacronym{mcmc}{MCMC}{Markov chain Monte Carlo}
\newacronym{ps}{PS}{point source}
\newacronym{kid}{KID}{kinetic inductance detector}
\newacronym{act}{ACT}{Atacama cosmology telescope}
\newacronym{cmb}{CMB}{cosmic microwave background}
\newacronym{hse}{HSE}{hydrostatic equilibrium}
\newacronym{los}{LoS}{line of sight}
\defcitealias{upp}{A10}
\defcitealias{hass}{H13}
\section{Introduction}
\label{intro}
  
The formation and properties of galaxy clusters provide invaluable insights into the structure and evolution of the Universe. These clusters, the largest gravitationally bound structures, arise from the gravitational collapse of matter and the accretion of material at the intersections of cosmic filaments \citep{cweb2,cweb1}. The imprint of structure formation is carried by the galaxy cluster number density and spatial distributions of galaxy clusters. Consequently, by studying their abundance across mass and redshift, we gain crucial information about the processes driving large-scale structure formation, offering a direct window into cosmology (for a review see e.g. \citealt{allen}). 

Recent advancements in observational techniques of the large-scale structure have significantly enhanced our understanding of cosmological parameters and structure formation mechanisms \citep{Huterer}. In the case of cluster cosmology, large galaxy cluster catalogues \citep{p27,sptbleem,act21,erass} have recently been used for cosmological studies using cluster counts \citep{p24,boc,erocc}. However, in the past, these results have been at odds with cosmological results from the CMB power spectrum analysis \citep{flatu,planckov}. Cluster-based cosmology relies primarily on the galaxy cluster mass estimation accuracy and challenges persist in doing so, primarily due to the indirect nature of mass measurement methods. Observations in X-ray and millimetre wavelengths can be used to estimate the cluster mass under hydrostatic equilibrium \citep{krav} and cluster gravitational lensing also serves as a tool for mass estimation but, these methods rely on assumptions about the physical properties of the cluster and its environment. Alternatively, cluster masses can be inferred from previously calibrated observable-mass scaling laws using observable mass proxies such as the X-ray luminosity and the thermal Sunyaev-Zeldovich effect (tSZ). Efforts to calibrate such scaling relations face hurdles like bias induced by non-gravitational processes (see \citealt{gwp_review} for a review of how mass proxies are affected by various biases). Despite these challenges, advancements in instrumentation, such as the New IRAM KIDs Array 2 (NIKA2), provide promising avenues for high angular resolution SZ observations of galaxy clusters. Successfully installed and commissioned at the Institut de Radio Astronomie Millimetrique (IRAM) 30-m telescope, NIKA2 is capable of detailed mapping of the ICM in high-redshift clusters,  offering insights into their thermodynamic properties and consequent variations in mass-observable scaling relations. With a resolution of $\sim 12^{\prime \prime}$ and $\sim 18^{\prime \prime}$ at $260 \, \mathrm{GHz}$ and $150 \, \mathrm{GHz}$, respectively  \citep{nika2,np}, it provides an angular resolution similar to the one reached by current X-ray observatories such as \textit{XMM-Newton} in terms of substructure resolution and ICM spatial extension \citep{upp_cc}. Furthermore, its dual-band feature enables a limitation of the systematics affecting single-frequency SZ mapping. For instance, it may be used to study contamination of the tSZ signal by dusty and radio point sources (see \citealt{adam16,ruppin}).

By combining multi-wavelength observations, including X-ray and optical data besides the tSZ data from NIKA2, with advanced analytical techniques, we aim to refine our understanding of galaxy cluster properties and their implications for the estimation of the cluster pressure and mass profiles. This holistic approach allows us to overcome the challenges associated with mass estimation and scaling relations, paving the way for more accurate cosmological models and a deeper understanding of the evolution of the Universe. In particular, we present the study for one of the observed clusters of the NIKA2 Sunyaev-Zeldovich Large Programme \citep[LPSZ,][see Section \ref{lpsz}]{lpszold,lpsz2}, ACT-CL J0240.0+0116. This cluster has one of the lowest masses and lowest redshifts in the sample, with the estimated cluster mass at overdensity $\Delta=500$\footnote{$\Delta = \left[M_{\Delta}/(\frac{4}{3}\pi r_{\Delta}^3)\right]/\rho_{crit}(z)$; where $\rho_{crit}(z)$ is the critical density of the Universe at redshift $z$} being $M_{\rm 500} = 3.3 \pm 0.8 \times 10^{14} \, h_{70}^{-1} M_{\odot}$ \footnote{The parameter $h_{70}$ is defined as $h_{70} = H_0/(70 \, \mathrm{km \, s^{-1}\, Mpc^{-1}})$; where $H_0$ is the Hubble constant. In \cite{hass}: $H_0 =
70\,  h_{70} \, \mathrm{km \, s^{-1}\, Mpc^{-1}}$; hence $h_{70} = 1$} and $z = 0.62\pm0.03$ as per the Atacama Cosmology Telescope (ACT) survey estimates in \citet[][hereafter H13]{hass}\footnote{We use the photometric redshift estimated in the ACT survey because the spectroscopic values were not available at the time that the LPSZ was established}. The NIKA2 maps of the cluster, due to their higher angular resolution than ACT, which has an arcminute level resolution, reveal strong contamination by a central point source that has a compensating effect on the observed negative tSZ signal of the cluster ICM. While a cluster with a similar (low) mass at higher redshift is easier to observe using NIKA2 because more signal fits into one telescope beam at $150 \, {\rm GHz}$, this target combines the challenges associated with the ICM study using the SZ effect, of a faint low-mass, low-redshift (within the LPSZ) cluster, contaminated by point sources, in characterising its dynamical state and recovering its pressure profile. We also estimate the cluster hydrostatic equilibrium (HSE) mass by combining the pressure profile from the SZ data and the electron density profile from the X-ray. Consequently, any bias incurred in estimating the pressure profiles is propagated in the recovery of the HSE mass. This work adds to the previous NIKA2 LPSZ cluster case studies, where the pilot study on PSZ2 G144.83+25.11 comprised a science verification study and demonstrated the ability of NIKA2 to detect substructures and its impact on ICM profile reconstruction \citep{ruppin}; the second study based on ACT-CL J0215.4+0030 as the worst-case scenario for the NIKA2 LPSZ in terms of mass and redshift \citep{keruzore}; and the study of CL J1226.9+3332, which explored the systematic effects in mass recovery and the mass bias incurred therein \citep{mirenlpsz}.

This paper is structured as follows. In Section \ref{clus}, we describe the target cluster being studied in this work, stating the information in the literature and building a case of interest for it to be analyzed using NIKA2 data. In Section \ref{multi}, we describe the data available for cluster ACT-CL J0240.0+0116, and how they are reduced and used to create a multi-wavelength composite image of the cluster. In particular, we describe the LPSZ and the NIKA2 data in Section \ref{nk2}, X-ray data in Section \ref{xmm}, and the optical data in Section \ref{sdss}. The process of recovery of the pressure and mass profiles of the cluster and the impact of the contaminating point source on these profiles are discussed in Section \ref{panco2} and Section \ref{masseval}, respectively. Section \ref{ds} analyses the cluster dynamical state by estimation of varied morphological parameters and the impact that the inferred dynamical state will have on the estimated mass. Finally, in Section \ref{actdiscussion}, we provide the discussion of our results and our conclusions.

Throughout this work, we assume a flat ${\rm \Lambda CDM}$ cosmological model. The adopted parameter values are $H_0 = 70\, {\rm km\, s^{-1}\, Mpc^{-1}}$, $\Omega_{\rm m,0} = 0.3$, and $\Omega_{\Lambda,0} = 0.7$. Under the assumption of this model, for our cluster redshift ($\sim 0.62$), an angular size of 1 arcminute corresponds to $407\, {\rm kpc}$.
\section{Cluster ACT-CL J0240.0+0116}
\label{clus}

This galaxy cluster, initially identified as WHL J024001.7+011606 using Sloan Digital Sky Survey (SDSS)-III data \citep{actsdss, aihara}, was recognised by detecting a high galaxy density at a specific redshift. It was later observed in SZ by the ACT survey \citepalias{hass} and renamed ACT-CL J0240.0+0116 (hereafter ACTJ0240). While included in the ACT SZ cluster catalogue \citep{act21} and observed in polarised mode \citep{actpol}, no detailed high-resolution SZ studies of the ICM have been conducted for this cluster. However, its ACT mass estimate was used to calibrate the optical mass proxy in a prior study (see e.g. \citealt{actopmass}). A summary of its position, mass, and redshift estimates from literature is provided in Table \ref{actsummary}.
\begin{table*}
\centering
\begin{tabular}{|ccccc|}
\hline

Parameter & Parameter description  & Value  & Wavelength & Reference \\
\hline
\hline 
 - & Alternate ID & WHL J024001.7+011606 & Optical & 1 \\

\hline

 (RA, Dec)  & Brightest central & $ (40.0072, 1.2685) $ & Optical & 1\\
 (Deg)  & galaxy position & $  $ &  & \\

   & Peak position of & $(40.0102, 1.2693) $ & Millimetre & 2\\
   & the SZ detection &  & &\\
\hline
$z$ & Photometric redshift & $0.5870$ & Optical & 1 \\
 & Photometric redshift & $0.62 \pm 0.03$ & Optical & 2 \\
 & Spectroscopic redshift & $0.6035$ & Optical & 3 \\
 & Spectroscopic redshift & $0.6041$ & Optical & 4 \\

\hline
 $M_{500}$ & Compton parameter-scaled & $3.3 \pm 0.8 $ & Millimetre & 2 \\
 $(10^{14}  \, \mathrm{M_{\odot}})$ & mass, estimated for a UPP$^{*}$ &  &  & \\

 & Compton parameter-scaled & $3.707 \substack{+0.639\\-0.545}$ &  Millimetre & 3 \\
  & mass, estimated for a UPP &  &  & \\

 & Optical richness & $3.65 \pm 1.12$ &  Optical & 4 \\
 & mass proxy &  &  & \\

\hline

\end{tabular}
\caption{A (non-exhaustive) summary of various properties of ACTJ0240, found in the literature. Coordinates are reported in the J2000 standard equinox. References: (1) \cite{actsdss}; (2) \citetalias{hass}; (3) \cite{act21}; (4) \cite{actopmass}. *UPP refers to Universal pressure profile from \citetalias{upp}.}
\label{actsummary}
\end{table*}

ACTJ0240 was observed by \textit{XMM-Newton} as part of the X-ray follow-up of 15 low-mass LPSZ clusters (details in Section \ref{xmm}). While the X-ray data were previously used to study the cluster's morphology and dynamical state, they have not been employed to infer ICM profiles. In \cite{actds}, ACTJ0240 was classified as disturbed using its X-ray images. Additionally, the cluster was included in a MeerKAT Exploration of Relics,
Giant Halos, and Extragalactic Radio Sources (MERGHERS) pilot study due to multi-wavelength indications of disturbance, though no diffuse radio emission was detected \citep{actmeerkat}, making another indication for the case that low-mass mergers may have ultra-steep radio spectra which makes the detection of a diffuse radio emission much more difficult than in massive mergers \citep{lowmassmergers}.

With the lack of previous ICM studies based on high angular
resolution SZ data and the establishment of the cluster being disturbed, the aforementioned studies pose ACTJ0240 as an interesting target for the NIKA2 LPSZ in terms of requiring careful
evaluation of the ICM, that investigates the disturbed state of the
cluster. The sensitivity and high resolution of NIKA2 enable us
to map the cluster ICM in detail and infer its dynamical state and
how the state can impact the inference of the cluster properties.
In this context, for this work, we analyse the NIKA2 SZ data of
this cluster, in combination with its X-ray and optical data.

\section{Multi-wavelength observations}
\label{multi}
\subsection{SZ observations}
\label{nk2}
\subsubsection{NIKA2 Sunyaev-Zeldovich Large Program (LPSZ)}
\label{lpsz}
The SZ data used in this work is obtained as a part of the NIKA2 LPSZ. The NIKA2 camera, installed at the IRAM 30-meter telescope in Pico Veleta (Spain), is a dual-band millimetre continuum instrument capable of mapping the sky at $150 \ \mathrm{GHz}$ and $260 \ \mathrm{GHz}$ using kinetic inductance detectors (KIDs) \citep{nika2,kids}. With a $6.5^{\prime}$ field of view and high angular resolution of $17.6^{\prime \prime}$ at $150 \ \mathrm{GHz}$ and $11.1^{\prime \prime}$ at $260 \ \mathrm{GHz}$, it offers sensitivities of $9 \ \mathrm{mJy \ s^{\frac{1}{2}}}$ and $30 \ \mathrm{mJy \ s^{\frac{1}{2}}}$, respectively, making it highly effective for SZ observations \citep{np}. The LPSZ, a guaranteed-time program of NIKA2, allocated (the now completed) 300 hours to observe the tSZ effect in 38 galaxy clusters, covering a mass range of $3 \leqslant M_{500} / 10^{14} \ \mathrm{M}_{\odot} \leqslant 11$ and redshifts of $0.5 < z < 0.9$. This high-resolution follow-up study, based on Planck \citep[PSZ2,][]{p27} and ACT catalogues \citep{act21}, aims to enhance SZ-based cosmological research \citep{lpsz2}.
\subsubsection{NIKA2 data}
ACTJ0240 was observed as a part of the LPSZ (project 199-16 of the NIKA2 guaranteed time) for a total integration time of $\sim 11 \ \mathrm{hours}$. The observation was centred at the X-ray peak estimated using \textit{XMM-Newton} data of the cluster, (R.A., Dec.)$_{J2000}$ = (02h40m03.384s, +01d15m58.32s) and was split over multiple sessions in the January and November 2020 observational shifts. The data, acquired simultaneously at 150 GHz and 260 GHz, comprises several raster scans following the pattern described in \cite{keruzore}. During the observations, sky dip readings were performed to measure the atmospheric opacity, required to account for the atmospheric attenuation. With a mean elevation of $\sim 51 ^{\circ}$, the average opacity at $225 \, {\rm GHz}$ for the observation was $\tau _{225} = 0.19 \pm 0.06$.

\subsubsection{Decorrelation process}
\label{decorr}

To calibrate and reduce time-ordered scan data into SZ images, we followed the baseline method used for NIKA2 performance assessment \citep{np}, which has since been adapted for other LPSZ studies \citep{ruppin,keruzore,mirenlpsz}. For this process, called ``decorrelation'', which removes noise due to the electronic components of the telescope and due to the atmosphere, we use the ``common mode one block'' or ``most correlated pixels'' method, detailed in \cite{np}. To prevent signal loss, the cluster region and point sources are masked. For ACTJ0240, we used a mask with a radius of $0.55 \, \theta _{500}$, where $\theta _{500} = 2.022^\prime$ based on ACT survey data. The $0.55 \, \theta_{500}$ mask size was optimised for the complete LPSZ sample to minimise signal loss while effectively removing noise. Point sources detected in the NIKA2 SZ maps with $\mathrm{S/N}>4$ are also masked.

\begin{figure*}[!ht]
   \centering
   \includegraphics[width=0.95\textwidth]{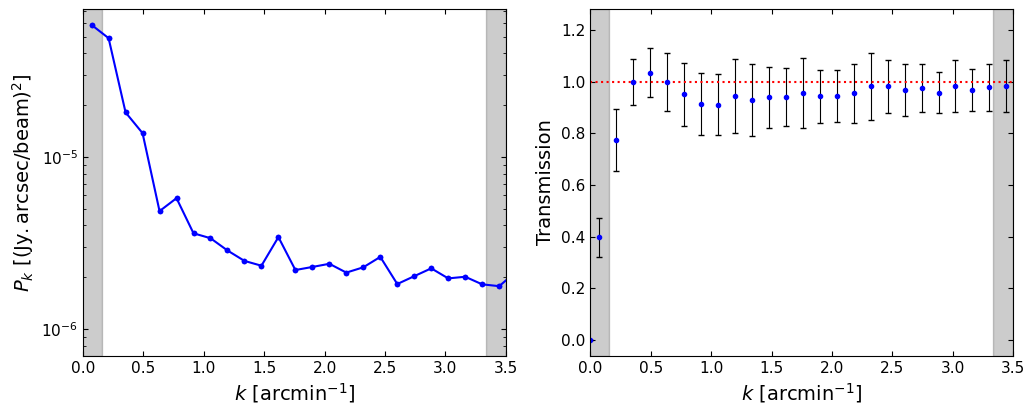}

   \caption{\textit{Left:} power spectrum of the residual noise at $150 \ \mathrm{GHz}$, estimated for ACTJ0240 using the half-difference maps. \textit{Right:} $150 \ \mathrm{GHz}$ pipeline transfer function for ACTJ0240. The shaded region on the left (right) in both plots depicts the NIKA2 field of view (instrumental FWHM).}
   \label{tfps}
    \end{figure*}

Despite decorrelation, some noise remains, and signal filtering outside the masked region occurs. Residual noise is assessed via the angular noise power spectrum obtained from null maps (jackknife maps, see \citealt{keruzore} for details), while the pipeline transfer function gauges signal filtering outside the mask (see \citealt{mirenlpsz} for the estimation process). For ACTJ0240, the left panel of Fig. \ref{tfps} shows the $150 \ \mathrm{GHz}$ noise power spectrum, indicating strongly correlated noise at large scales. The right panel shows the transfer function at $150 \ \mathrm{GHz}$, revealing signal suppression at large scales ($k < 0.5 \text{ arcmin}^{-1}$) but good preservation at smaller scales. These two quantities are utilised to estimate thermodynamic properties from the SZ map, as discussed in Section \ref{panco2}.

\subsubsection{NIKA2 tSZ maps}
\begin{figure*}
   \centering
   \includegraphics[width=0.95\textwidth]{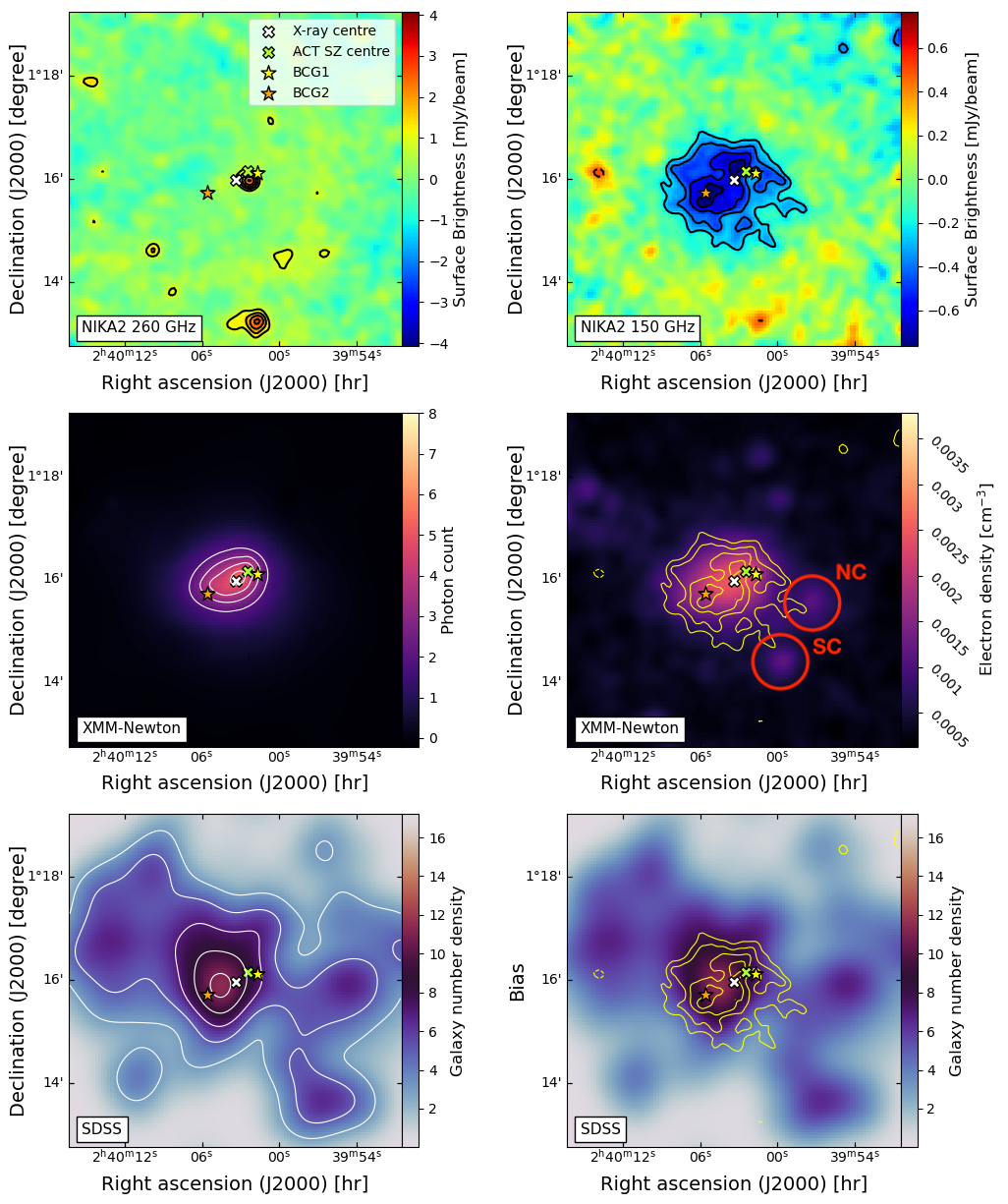}

   \caption{Multi-wavelength view of ACTJ0240. The images are square grids of size $6.5^{\prime} \times 6.5^{\prime}$, centred at the X-ray peak of the cluster. In all the images, we show the ACT SZ peak (green cross), the \textit{XMM-Newton} image peak (white cross), and the two cluster BCGs (yellow and orange stars). \textit{\textbf{Top panels:}} the $260 \ \mathrm{GHz}$ (left) and $150 \ \mathrm{GHz}$ (right) NIKA2 surface brightness images. The black contours in both images show the SNR levels starting from $\pm 3 \sigma$, spaced by $2 \sigma$. For visualisation, the left (right) map has been smoothed with a Gaussian kernel of $6^{\prime \prime}$ ($10^{\prime \prime}$). \textit{\textbf{Central panels:}} On the left is the wavelet-smoothed \textit{XMM-Newton} photon count image of the cluster with the white contours depicting signal above $2 \sigma$ and each successive contours are spaced by $1 \sigma$, produced using the technique detailed in \cite{bourdin08}. The right side shows the X-ray inferred electron density map (smoothed using a $7.5^{\prime \prime}$ Gaussian kernel), overlaid with SZ contours corresponding to the top-right panel. \textit{\textbf{Bottom panels:}} Both the images show the galaxy number density of the cluster estimated from SDSS data. The image on the left is overlain with white isodensity contours and the yellow contours in the right-hand side image are the same as the top-right panel.}
              \label{actnk}%
    \end{figure*}

The final surface brightness images from the NIKA2 observations of ACTJ0240 are shown in the top panel of Fig. \ref{actnk}. The $260 \ \text{GHz}$ map (top left) shows no positive thermal SZ effect signature, as expected, due to noise levels of NIKA2 exceeding the expected signal ($\sim 0.2 \ \mathrm{mJy/beam}$). However, a strong \acrshort{ps} near the cluster centre is visible. In the $150 \ \text{GHz}$ map (top right), the cluster appears as a negative signal (SZ effect), depicting a characteristic negative signature of the thermal SZ effect expected at this frequency, with a peak detection of $\mathrm{S/N}=9$ at the centre. A \acrshort{ps} near the centre, at position (R.A., Dec.)$_{J2000}$ = (02h40m02.4s, +01d15m58.2s), creates a "hole" in the map by compensating for the SZ effect decrement, potentially affecting ICM property estimation from this map. The SZ map also reveals two potential, interacting tail-like structures extending southwest, whose plausibility is confirmed by complementary observations in X-ray and optical, as discussed in the following sections. 

Additionally, Fig. \ref{actnk} highlights different cluster centres: the ACT SZ peak (green cross), the X-ray peak from \textit{XMM-Newton} (white cross), and the two brightest central galaxies (BCG1 and BCG2, marked by yellow and orange stars). For ACTJ0240, the difference in magnitude of the two brightest galaxies, both sitting at the same redshift as the cluster, is negligible (magnitude difference $\sim 0.038$ in the SDSS-III r band), rendering it difficult to define a clear BCG candidate that coincides with the centre of the cluster potential well. The NIKA2 $150 \, {\rm GHz}$ map displays two distinct SZ peaks that coincide with the BCGs (once the point source contamination is taken into account): a potential sign that there are two merging components in this cluster system.
    
\subsection{\textit{XMM-Newton} observation and image}

\label{xmm}
ACTJ0240 was observed by XMM-\textit{Newton} (Obs ID 0800971001) for a total observation time of 69 ks. The raw data were processed by following the procedure outlined in \cite{iacopotemp} using the Science Analysis System (SAS)
version $16$\footnote{https://www.cosmos.esa.int/web/xmm-newton/sas} and the embedded Extended Source Analysis System (ESAS, \citealt{snowden08,kuntz08}). Contamination induced by the high-energy particles was reduced retaining only events with the keyword \texttt{PATTERN} $<13$ and $<5$ for the European Photon Imaging Camera (EPIC) pn \citep{struder2001} and Metal Oxide Semiconductor (MOS 1 and 2, \citealt{turner2001}) cameras, respectively. Parts of the data involving an anomalous count rate were removed following the procedure described in \cite{prattim}. The useful exposure time after the cleaning procedures is 36/29 ks for the MOS/pn detectors. Further, the high-energy particles interacting with the detectors and with the telescope itself, generate events that contaminate the signal of interest. We used the filter wheel closed datasets (see \citealt{pratt10} for details) to evaluate and remove this component. The vignetting correction was applied to data following the process outlined in \cite{arnaud01}. Finally, point sources were removed from the analysis by running the Multiresolution wavelet software \citep{wlfilter} over exposure-corrected images in the $[0.3-2.0] \ \mathrm{keV}$ and $[2.0-5.0] \ \mathrm{keV}$ bands. Further inspection was done by eye to recognise missing sources/structures or remove fake detections. The sky foreground and background in X-ray data include a local component and an extragalactic component (see, \textit{e.g.}, \citealt{snowdenbg,kuntz,giacconi}). This sky component was modelled and consequently removed from the data, as a constant background in the surface brightness estimation process. The instrumental and sky background cleaned data was projected to a common sky projection to obtain a combined (MOS1, MOS2, and pn) image of the cluster. The wavelet-smoothed photon count image of ACTJ0240 is shown in the central-left panel of Fig. \ref{actnk}, which shows us the general alignment of the gas distribution depicted individually by the SZ and X-ray data.

\subsubsection{The X-ray ICM temperature and electron density}
\label{tempden}
The X-ray inferred electron density was determined following the methodology described in Sections 3.2 and 3.4 of \cite{iacopotemp}. Briefly, the vignetted-corrected and background-subtracted surface brightness profile was extracted from concentric annuli centred on the peak of X-ray emission. We then deprojected this profile employing the regularization technique outlined in \cite{croston} correcting for the \acrfull{psf}. The 3D temperature profile was determined following the procedure described in \cite{pratt10} and in Section 3.3 of \cite{iacopotemp}. We firstly extracted the spectrum from concentric annuli centred on the X-ray peak, with the bin width for each annulus defined such that the signal-to-noise ratio of at least $30\sigma$ was attained above the background level in the [0.3-2] keV band. The binning also ensured a minimum of 25 counts per energy bin after subtracting the instrumental background. The spectrum was fitted using a combination of models to include both the cluster signal and the sky/background emissions using XSPEC\footnote{https://heasarc.gsfc.nasa.gov/xanadu/xspec/} \citep{xspec,xspecman}. The deprojected 3D temperature profile, $T_{3D}$, is obtained following the technique described in Sect. 2.3.2 of \cite{bartalucci18}.

\begin{figure*}
   \centering
   \includegraphics[width=0.8\textwidth]{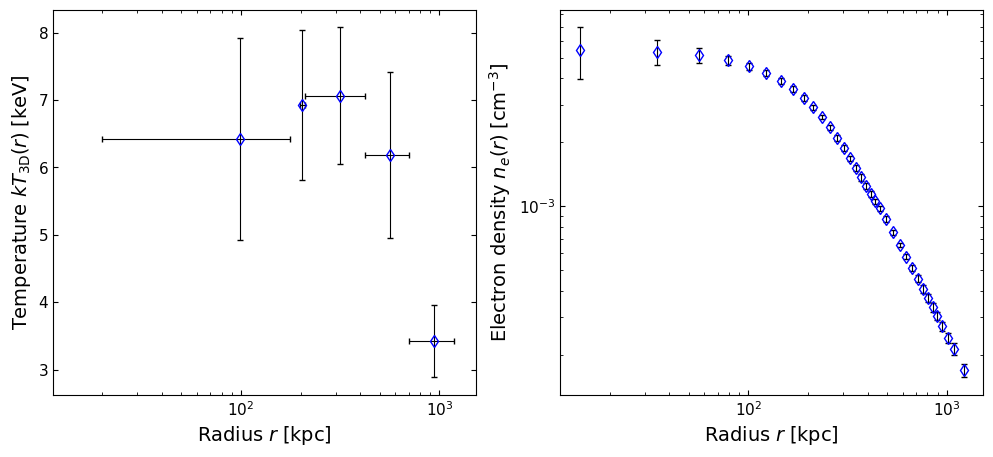}
   \caption{3D temperature (left) and electron density (right) profiles from \textit{XMM-Newton} observations of ACTJ0240. Vertical errors show $1\sigma$ uncertainties and horizontal error bars in the temperature profile indicate the radial binning used.}
\label{temprof}%
\end{figure*}

The X-ray-derived 3D temperature and electron density profiles are shown in the left and right panels of Fig. \ref{temprof}, respectively. The limitation of having to sample the X-ray temperature profile using broad radial bins to ensure enough statistics, for a high redshift cluster, is evident via the temperature profile. Despite this, the errors in the values of the X-ray temperature, depicted by the error bars in the left panel of Fig. \ref{temprof}, are high ($>10\%$). We notice that the density profile features a plateau in the cluster core. One potential explanation for this could be the absence of a cool cluster core, as probed by the X-ray observations \citep{ccproof}.

Finally, we produced the electron density map following the procedure described in Section 3.2 of \cite{remi17}. This is shown on the right side of the central panel of Fig. \ref{actnk}. Of particular interest, in terms of the morphological study of the cluster, are the two gas clumps (labelled `NC' and `SC') we notice right next to the extended tail-like structures detected in the southwest region of the SZ map (shown in the middle-right panel of Fig. \ref{actnk} using yellow contours). We discuss them in detail in Section \ref{ds}.

\subsection{SDSS Optical data and image}
\label{sdss}
We utilised information on galaxies obtained from the SDSS data release 16 \citep{sdssr16} to create an optical galaxy number density map for ACTJ0240. Unfortunately, the available spectral data for galaxies in the cluster region and redshift were insufficient for a reliable estimation of their velocity dispersion and, consequently, the dynamical mass of the cluster. As an alternative, we employed the photometric data of galaxies within the cluster region, specifically utilizing information from the $g$, $r$, and $i$ bands. To identify member galaxies, red-sequencing (refer to \textit{e.g.} \citealt{rs1} for an early reference) was applied to the data. For ACTJ0240, due to its high redshift, we used the magnitude in the r band and colour using the r and i bands to find the red sequence of the member galaxies. The identified member galaxies were used to create a galaxy number density map of the cluster. To estimate the number density map, we smoothed the galaxy map in the region of interest with a Gaussian kernel whose normalisation was equal to the total number of galaxies present in the region. This provided the number density of galaxies at each point in the grid. The resulting number density map can be seen in the bottom panel of Fig. \ref{actnk}. On the left side, we show the number density map and on the right, we show the map together with the SZ map contours (in yellow). From the left figure, we can see that one of the BCGs (BCG2, shown using an orange star) is closer to the peak of the galaxy distribution while the other (BCG1, shown using a yellow star) is far from the peak of the galaxy distribution. This is unusual since, for a relaxed or virialised system, the brightest galaxies sit at the centre of the halo. We discuss in Section \ref{ds} how the difference in magnitude can also indicate the dynamical state of the cluster.
\section{Estimation of the ICM pressure profile}
\label{panco2}
The previous sections described the creation and visualization of multi-wavelength images of galaxy cluster ACTJ0240. To better understand the cluster in the LPSZ context, it is crucial to characterise the ICM. One key LPSZ goal is to estimate and calibrate cluster masses to explore the $Y-M$ scaling law at medium to high redshifts \citep{lpsz2,alice}. We estimate the pressure profile, and consequently the cluster mass, from the NIKA2 SZ map at $150 \, {\rm GHz}$ using the \texttt{PANCO2} software, which uses a forward modelling approach to estimate the pressure profile from a SZ map \citep{pc2}.

\subsection{Reconstruction of pressure profile from NIKA2 data}
The SZ effect distortion in the CMB, quantified by the Compton parameter ($y$), is proportional to the electron pressure $P_e$ of the ICM integrated along the \acrfull{los}:
\begin{equation}
    y(\hat{n}) = \frac{\sigma_T}{m_e c^2} \int P_e(\hat{n},l) \, \mathrm{d}l.
    \label{y-p}
\end{equation}
Here, $m_e$, $c$, and $\sigma_T$ represent the electron mass, speed of light, and Thomson scattering cross-section, respectively. Assuming spherical symmetry, we render the pressure distribution to depend only on the distance from the cluster centre. Under these circumstances, using Abel transformation, Eq. (\ref{y-p}) can be projected into a 2D radial profile. To fit such a projected profile to our tSZ map, we initially describe our 3D pressure profile using the so-called `radially-binned' model also referred to as the `non-parametric' (NP) model (in \citealt{ruppinnp,nonparam}), where the pressure profile is described by a radially falling power-law function distributed in concentric spherical shells, given by:
\begin{equation}
    P_e\left(r_i<r<r_{i+1}\right)=P_i\left(\frac{r}{r_i}\right)^{-\alpha_i},
    \label{npform}
\end{equation}
here $P_i$ and $\alpha _i$ are the values of the pressure and the numerical slope for the radial bin positioned at radius $r_i$. The functional form described provides a less model-dependent method for estimating the pressure profile, accommodating irregularities like jumps or shocks. To obtain a projected map corresponding to the 3D model of the above pressure profile, the model corresponding to Eq. \ref{npform} is first integrated along the LoS to obtain a Compton parameter map ($y$ map) in the cluster region. These analytical integrals have been calculated as in \cite{nonparam}. This $y$ map is then converted to an SZ surface brightness map using a telescope calibration factor (see \citealt{np}). The calibration factor, however, depends on the effective bandwidths and beam of NIKA2 and the shape of the SZ effect spectrum and consequently the ICM temperature. To allow the variation in calibration factor based on these parameters, \texttt{PANCO2} treats it as a nuisance parameter in the likelihood, with a Gaussian prior and initial guess of $-11.9$ at $150 \, {\rm GHz}$ \citep{np,ruppin}. The obtained SZ map then goes through two forms of filtering: i) convolution with the instrumental beam characterised by a Gaussian filter of FWHM of the $150 \ \text{GHz}$ channel ($18 ^{\prime \prime}$); ii) the filtering resulting from the NIKA2 decorrelation method, which is quantified by the analysis transfer function described in Section \ref{decorr}. The resulting map is used as a model to compare to the surface brightness map of the SZ effect observed by NIKA2 together with the consideration of a zero level (additive bias), which is fitted as a nuisance parameter. 

The hydrostatic mass associated with this pressure profile is estimated under the assumption of HSE of the cluster medium \citep{krav} along with spherical symmetry. Under this circumstance, the mass enclosed with a spherical volume of radius $r$ is given by:
\begin{equation}
    \label{hsemass}
    M_{\mathrm{HSE}}(<r)=-\frac{r^2}{G \mu m_{\mathrm{p}} \ n_e(r)}\frac{\mathrm{d} P_{\mathrm{e}}(r)}{\mathrm{d} r},
\end{equation} where $G$, $\mu$, $m_p$, and $n_e(r)$ are the universal gravitational constant, mean molecular weight of the ICM
gas, the proton mass, and the ICM electron density (at radius $r$),  respectively. Following \cite{ettorimu}, we consider $\mu = 0.6$ to be a good approximation for the ICM gas.
The NP model is very sensitive to discontinuities in the first derivative of the pressure profile which, following Eq. \ref{hsemass}, would then lead to negative, non-physical results in the mass profile derived under the HSE assumption. To prevent this from happening we fit a generalised Navarro-Frenk-White (gNFW, \citealt{nagai}) model to the best-fit values of the NP pressure profile data points obtained above. The gNFW model is given by:
\begin{equation}
    \frac{P_e(r)}{P_{500}}=\frac{P_0}{\left(\frac{r}{r_s}\right)^\gamma\left[1+\left(\frac{r}{r_s}\right)^\alpha \right]^{\frac{\beta-\gamma}{\alpha}}},
    \label{gnfw}
\end{equation}
where $P_0$ and $P_{500}$ are the normalization constant and pressure given by Eq. 5 of \citet[][hereafter A10]{upp}. Parameters $\beta$ and $\gamma$ represent the external and internal slopes of the pressure profile, with $r_s$ marking the characteristic radius at which the profile slope changes and parameter $\alpha$ quantifies the sharpness of this transition. 

\subsubsection{Point source contamination} 
When estimating the pressure profile, addressing the central contaminating PS (Section \ref{nk2}, Fig. \ref{actnk}) is crucial due to its potential biasing effect on the estimated pressure profile. To emphasise this point, we adopt three approaches to handling this \acrshort{ps}: i) Be blind to the PS: completely ignore its presence to see the impact on ICM analysis; ii) Masking the PS: in this case, we assign zero weight to the pixels at the position of the PS in both, the NIKA2 SZ map and model map during the likelihood analysis; iii) Modelling the PS: \texttt{PANCO2} offers the possibility of treating PS contamination as a nuisance parameter in the ICM pressure profile fitting \citep{pc2}. Here, the source flux is then considered as an additional parameter of the model. Previously mentioned LPSZ studies used flux estimates from catalogues like PACS \citep{pacs}, SPIRE \citep{spire}, and FIRST \citep{firstcat} to determine spectral energy distributions and flux priors for PSs in their models. Unfortunately, our PS is not present in any external submillimeter or radio catalogues. Our PS is unlikely to be a radio-loud active galactic nucleus since these are typically centred at the BCG and our source is significantly displaced from the BCG centre. Additionally, a radio source would likely overshadow the cluster negative tSZ peak for our low-mass cluster. We therefore consider the PS to be a submillimeter galaxy, which is a common contaminant in SZ observations. For such galaxies, the typical flux ratio between the NIKA2 bands is $S_{\text{150}}/S_{\text{260}}=0.2$ \citep{ricci}. Thus, we estimate the PS flux in the $260 \ \text{GHz}$ band and set an upper limit for the $150 \ \text{GHz}$ band at $0.3$ times this value.

\subsection{Markov Chain Monte Carlo Fitting}

We use an MCMC algorithm to fit our models to the data and find the posterior distribution of the parameters ($\vartheta$) given the NIKA2 measurements. This posterior distribution, $P(\vartheta \mid D)$, is defined by Bayes' theorem, where $D$ refers to the NIKA2 $150 , {\rm GHz}$ surface brightness map and the $Y_{\text{500}}$ measured by ACT \citepalias{hass}. We model the prior distribution as the product of independent Gaussian distributions centred on universal pressure profile (UPP) values from \citepalias{upp}, with a standard deviation of five times the central values. The likelihood function, $L(D \mid \vartheta)$, is given by a multivariate Gaussian function:
\begin{equation}
    \begin{split}
        \log L(D \mid \vartheta)=-\frac{1}{2} \sum_{i=1}^{n_{\text {pixels }}}\left[(\mathcal{M}(\vartheta)-D)^T C_{\text {pix-pix }}^{-1}(\mathcal{M}(\vartheta)-D)\right]_i \\ -\frac{1}{2}\left(\frac{Y_{500}(\vartheta)-Y_{500}^{A C T}}{\Delta Y_{500}^{A C T}}\right)^2.
    \label{likelihood}
    \end{split}
\end{equation} 
Here, $Y_{500}^{\text{ACT}}$ and $Y_{500}(\vartheta)$ are the ACT-measured and model-integrated SZ signals, respectively, and $\Delta Y_{500}^{\text{ACT}}$ is the uncertainty in the ACT measurement. Including this information in the likelihood function ensures that the information on large scales, filtered by the LPSZ decorrelation process, can be captured via the integrated SZ signal of the cluster that is reliably measured in surveys with a larger field of view (ACT in this case). The model map is represented by the term $\mathcal{M}(\vartheta)$ and $C_{\text {pix-pix }}^{-1}$ represents the inverse pixel-to-pixel covariance matrix of the noise in the NIKA2 map. This matrix can be calculated from the power spectrum we earlier estimated (see e.g. \citealt{keruzore} for details).

For the gNFW model ($P_{\text{gNFW,NP}}$) fit to the binned NP profile ($P_{\text{NP}}$), the likelihood function is given by:
\begin{equation}
    \label{likelihoodmass}
    {\rm log} \, L(\vartheta)=-\frac{1}{2}\left[(P_{\text{gNFW,NP}}(\vartheta)-P_{\text{NP}})^T C^{-1}(P_{\text{gNFW,NP}}(\vartheta)-P_{\text{NP}})\right].
\end{equation}
The covariance matrix for data points $P_{\text{NP}}$ is given by $C$. The parameter vector $\vartheta$ is constructed by the gNFW model parameters described in Eq. (\ref{gnfw}). We ensure that the mass profile is non-negative and monotonically increasing to address potential violations of the HSE assumption due to a disturbed cluster state and a consequent non-physical HSE mass profile. Accordingly, following Eq. \ref{hsemass}, we impose the conditions $\frac{dP_e(r)}{dr} < 0$ and $\frac{dM_{\rm HSE}(r)}{dr} > 0$ during the fitting process corresponding to Eq. \ref{likelihoodmass}.

The MCMC implementation in this work uses Python \texttt{emcee} library \citep{emcee}, with burn-in phase, chain adaptation, autocorrelation testing, and the Gelman-Rubin convergence criterion \citep{gelman} adapted as $\hat{R} < 1.02$, used for convergence tests.
For fitting the NP model to the SZ maps, we sampled the pressure profile in five logarithmically-spaced bins between the NIKA2 beam and FoV, centred at the X-ray peak, using $30$ chains, discarding $1000$ burn-in steps. We report the best-fit profiles as the median and related uncertainty (16th and 84th quantiles) of the posterior distributions. The pressure profiles from the SZ map with the NP model are referred to as $P_{\text{NP}}(r)$, and the gNFW fit to NP data points as $P_{\text{gNFW,NP}}(r)$. If the PS is ignored, the gNFW fit does not converge even with $5\times 10^{5}$ iterations, as the PS creates a dip in the pressure profiles not captured by the gNFW model. For the two other cases, we use the same radial range described above and fit the gNFW model on the NP point using 50 chains and 1500 burn-in iterations.

\subsection{Pressure profiles}
\begin{figure}
   \centering
   \includegraphics[width=\columnwidth]{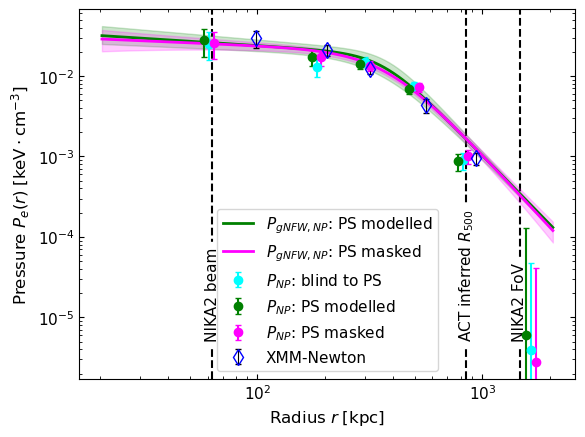}
      \caption{Pressure profiles of the ICM of ACTJ0240. Dotted markers show results from modelling the NIKA2 $150 \ \text{GHz}$ SZ maps using an NP pressure profile, with $1\sigma$ errors. The points are slightly horizontally displaced for visual clarity. Solid lines represent the best-fit gNFW model to the binned NP data, with shaded regions indicating the 16th and 84th percentiles. Blue empty diamonds show the pressure profile obtained from the \textit{XMM-Newton}. Vertical dashed lines indicate the NIKA2 beam, $R_{500}$ from the ACT survey, and the NIKA2 FoV, from left to right.}
              \label{pressureprof}%
    \end{figure}

The inferred median pressure profiles and their uncertainties are shown in Fig. \ref{pressureprof}. The best-fit NP pressure profiles, $P_{\text{NP}}(r)$, are marked with circles: cyan for ignoring the PS, magenta for masking it, and green for modelling it as a nuisance parameter. Profiles are consistent within their $1\sigma$ error bars, though ignoring the PS (cyan) shows lower pressure at $\sim 200 \ \mathrm{kpc}$, as the PS compensated, negative SZ signal is not accounted for. The gNFW fits to the NP data points, $P_{\text{gNFW,NP}}(r)$, are shown with solid curves in the same colours. Consistency is found between masking and modelling the PS, especially at $R_{500}$. However, $P_{\text{NP}}$ and $P_{\text{gNFW,NP}}$ agree only in the core and inner regions ($r \lesssim 800 \ \mathrm{kpc}$), with significant deviations around $R_{500}$. Note that this discrepancy may result from the conditions imposed to ensure a non-negative, monotonically increasing mass profile when applying HSE in a seemingly disturbed cluster. These constraints significantly limit the first and second derivatives of, and hence prevent the fitted $P_{\text{gNFW,NP}}(r)$ profiles to consistently agree with the $P_{\text{NP}}$ data points, consequently impacting the derived HSE mass.

The X-ray pressure profile from \textit{XMM-Newton} (blue diamonds) matches the $P_{\text{gNFW,NP}}(r)$ profiles (solid curves) within errors, reiterating that SZ data can infer ICM properties as well as X-ray data, which has a redshift-dependant emission. The SZ profiles $P_{\text{NP}}(r)$ (dots) agree with the X-ray profile for $r \lesssim 800 \ \mathrm{kpc}$. All estimates are consistent in this range, regardless of PS treatment or the wavelength used. The only discrepancy is at $r \sim 200 \ \mathrm{kpc}$ when ignoring the PS (cyan points), where the SZ signal compensation by the PS leads to an underestimated pressure.

\section{Estimation of the cluster hydrostatic equilibrium mass}
\label{masseval}

\begin{figure}
   \centering
   \includegraphics[width=\columnwidth]{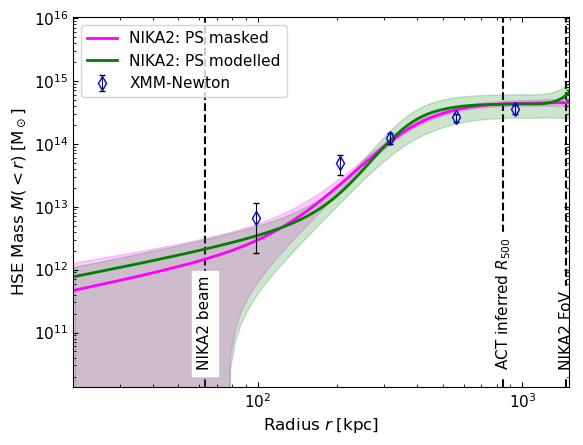}

   \caption{HSE mass profile estimates for ACTJ0240 inferred using NIKA2 and \textit{XMM-Newton} data. The solid magenta and green lines are mass profiles estimated by combining the X-ray density profiles (right panel Fig. \ref{temprof}) and the NIKA2 pressure profiles $P_{\mathrm{gNFW,NP}}$ (Fig. \ref{pressureprof}), in the scenarios where the \acrshort{ps} is masked, and modelled, respectively. The shaded areas depict the 16th and 84th percentiles of these estimates. Blue, empty markers correspond to the mass profile inferred solely using the X-ray data, with their $1 \sigma$ uncertainties.}
              \label{massprof}%
    \end{figure}

From the posterior distribution of $P_{\mathrm{gNFW,NP}}$ and the X-ray electron density profile $n_e$, which we interpolate to combine with the pressure profiles, we derived the HSE mass profile distribution using Eq. (\ref{hsemass}). We used the median and 16th and 84th percentiles of this distribution as the mass profile estimates and errors. The HSE mass profiles with PS masked and modelled, as well as the X-ray-only profile, are shown in Fig. \ref{massprof}. The two NIKA2 mass profiles (solid curves) are consistent with each other and with the X-ray-only profile (blue empty markers). The inferred mass profiles are used to calculate integrated quantities: mass $M(<r)$ and Compton parameter $Y(<r)$ within a given radius $R$. We define the radius for estimating these quantities by calculating radii corresponding to a density contrast $\Delta=500$ from the mass profile posterior distribution. We then estimate the posterior distributions for the integrated mass $M_{500} = M(R_{500})$ and the Compton parameter $Y_{500}$, where $ Y = 4\pi\sigma_T/m_ec^2 \int_{0}^{R} r^2 P(r) \,dr$. The median values and uncertainties for $R_{500}$, $Y_{500}$, and $M_{500}$ are reported in Table \ref{int_quant}. Note that these integrated quantities, while marginalised from the posterior distributions, are correlated. This correlation is assessed by calculating the covariance from the Markov chain samples, which is crucial for the $Y_{500} - M_{500}$ scaling relation in the LPSZ study \citep{alice}.

\begin{figure}
   \centering
   \includegraphics[width=\columnwidth]{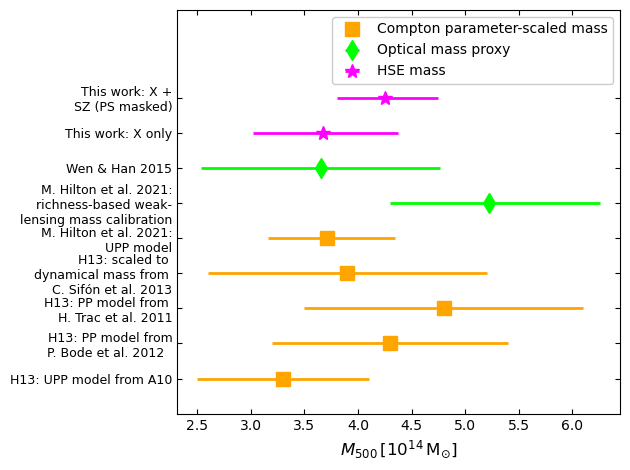}

   \caption{Mass estimates for ACTJ0240 in literature, compared to this work. The median mass estimate and the corresponding 16th and 84th percentile errors from this work are shown using the magenta-star marker. Square-orange markers represent the mass values estimated from the Compton parameter-mass scaling relations for different pressure profile (PP) models mentioned in the plot (references: \citetalias{hass,upp}; \citealt{bode,trac,sifon,act21}). Lime-diamond markers are the optical mass proxies calibrated through various sources mentioned in \cite{act21} and \cite{actopmass}.}
              \label{masscomp}%
    \end{figure}

In addition to $M_{500}$, we estimate $M_{1000}$ as $4.12 \substack{+0.50\\-0.40} \times 10^{14} \, \mathrm{M}{\odot}$ ($4.14 \substack{+1.75\\-1.14} \times 10^{14} \, \mathrm{M}{\odot}$) and $M_{2500}$ as $3.09 \substack{+0.54\\-0.51} \times 10^{14} \, \mathrm{M}{\odot}$ ($3.53 \substack{+1.72\\-1.42} \times 10^{14} \, \mathrm{M}{\odot}$) for the cases where the PS is masked (modelled). The overdensity $\Delta=1000$ shows the most stability to different PS treatments even though the integrated quantities are consistent across both methods, at all overdensities. However, uncertainties are lower when the PS is masked, likely due to the lack of external catalogue data and the poorly informed prior for the modelled PS preventing us from achieving a precise estimation of the
contamination. Thus, masking the PS results in more precise mass estimates and serves as our baseline estimate for comparisons. Fig. \ref{masscomp} shows a comparison of our mass estimates (from the combination of SZ and X-ray data and X-ray-only) with those in the literature. The X-ray-only mass, $M_{500} = 3.67\substack{+0.70\\-0.65} \times 10^{14} \,\mathrm{M}_{\odot}$, with $R_{500}$ inferred via the X-ray mass profile, is slightly lower than the SZ and X-ray combined mass ($4.25\substack{+0.50\\-0.45} \times 10^{14} \,\mathrm{M}_{\odot}$), but the two are consistent within their uncertainties. Our combined SZ, X-ray mass estimate is also consistent with the mass calibrated from the $y-M$ scaling laws in \citetalias{hass} and \cite{act21}, as well as the optical mass proxy values from \cite{actopmass}. In general, we show consistency with previous results, albeit, our precision is higher than any of these studies, highlighting the advantage of adding high-resolution NIKA2 data for improving the mass estimate uncertainty.

\begin{table}
\centering
\begin{tabular}{|c||ccc|}
\hline
Parameter & $M_{500}$  & $R_{500}$  & $D_A^2Y_{500} $ \\
 & $[10^{14} \mathrm{M}_{\odot}]$  & $[\mathrm{kpc}]$  & $[\mathrm{kpc}^2]$ \\

\hline \hline
Analysis &  & & \\
\hline 
 \acrshort{ps} modelled & $4.15\substack{+1.84\\-1.15}$  & $897.30\substack{+10.62\\-38.90}$  & $47.92\substack{+12.28\\-4.94}$  \\
PS masked & $4.25\substack{+0.50\\-0.45}$ & $907.63\substack{+37.52\\-26.66}$ & $52.28\substack{+7.48\\-5.64}$ \\

\hline
\end{tabular}
\caption{Median values of $M_{500}$ and $R_{500}$ estimated via the HSE analysis using SZ and X-ray data. We also report $Y_{500}$, obtained by integrating the inferred pressure profile. The errors are the 16th and 84th percentiles of parameter estimate distribution.}
\label{int_quant}
\end{table}

\section{Cluster dynamical state}
\label{ds}
The HSE assumption is valid for dynamically relaxed clusters but may bias mass estimates for disturbed clusters, potentially underestimating mass by up to 30\% \citep{giulia}. Thus, assessing the dynamical state of clusters is vital for accurate cosmological studies. Cluster morphology, invariably impacted by disturbances in the cluster, has often been studied as a key indicator of the dynamical state \citep{morpho_lovisari,lopes,luca,ricci_morpho,adam_morpho}.

For ACTJ0240, visualizations in Section \ref{multi} reveal signs of disturbance. Ideally, the SZ and X-ray centres should align with the BCG in a relaxed cluster \citep{actds,ota}. However, here, the X-ray and SZ centres are offset from the BCGs, indicating a misalignment of gas and galaxy distributions (see Fig. \ref{actnk}). Additionally, two bright galaxy candidates for the BCG are also displaced from each other, further suggesting a non-homogeneous galaxy distribution. In addition to the displacement of cluster components, we observe two potential tail-like structures in the SZ image (top right panel of Fig. \ref{actnk}), oriented northeast-southwest. The X-ray image (middle panel of Fig. \ref{actnk}) shows two clumps, labelled `NC' and `SC' corresponding to these SZ structures. The X-ray, sensitive to colder, clumped gas, allowed us to measure the temperature of the southern clump (SC) as $2.3\substack{+2.05\\-0.87} \, \mathrm{keV}$; however, the northern clump temperature was unconstrained. The pressure in these regions was below the NIKA2 noise level, so no SZ signal was visible for these clumps. Additionally, there is a higher galaxy density in the southwest near the SZ tails, trailing the X-ray clumps. The SZ tails, X-ray clumps, and trailing galaxies suggest in-falling structures, indicating that the cluster is disturbed and interacting. These features also make our cluster system resemble a low-mass version of the MACS J0717.5+3745 merger studied in \cite{adamnika}, with similar SZ peaks and associated BCGs and clump structures like `NC' and `SC' in our cluster.  Having said that, we can only speculate on this similarity since we do not perform a quantitative verification of this potential merger scenario.

\subsection{Morphological indicators}

Given the interaction and disturbance signs in the cluster, we analyze its ICM and morphology using high-resolution NIKA2 SZ images, X-ray, and optical data. This is the first NIKA2 LPSZ image analysis for this purpose. We apply eight morphological indicators (discussed below) to the SZ and X-ray image and estimate the magnitude gap ($\Delta m_{12}$) from the optical image to infer the dynamical state of ACTJ0240.

Disturbed clusters, often due to mergers or interactions, exhibit spherically asymmetrical ICM distributions. We measure this asymmetry, $A$, by evaluating the normalised differences between various rotations of the ICM map \citep{schade}. For $A$, we use the maximum difference from four rotations (at $90^\circ$, $180^\circ$, and mirroring the SZ and X-ray maps along their axes). Similarly, the strip $S$ \citep{c18} quantifies asymmetry by comparing $N$ light curves passing through the same centre but aligned at different angles. We extract four light curves, each rotated by $45^\circ$ increments. Mergers or turbulence in the ICM can also shift the centroid of the ICM signal when measured at various apertures from the cluster centre. We measure average centroid shifts $w$ \citep{mohr, Poole2006, Cassano2010} by analyzing ten concentric circular regions and comparing centroid positions with other tracers, such as BCG locations.

Multipole decomposition of the ICM SZ or X-ray signals helps detect cluster substructures \citep{powerr}. We use the third-order ratio $P_3/P_0$, which is sensitive to ICM bimodality.

The ratio ($c$) of surface brightness between two concentric regions, introduced by \citet{santos}, is a cost-effective indicator for detecting cool cores in X-ray cluster images without needing detailed spectroscopy. It is also widely used to identify more diffuse structures in disturbed clusters compared to relaxed ones.

Ultimately, we combine our indicators into a single metric $M$ following \citet{luca}, using {\sc The Three Hundred} \citep[hereafter The300]{300} parameters for mean and variance, to assess the cluster dynamical state. Additionally, we include the offsets of the SZ and X-ray centroids from the BCGs as dynamical state indicators. For the optical data, we use the magnitude gap ($\Delta m_{12}$, \citealt{maggap}), which measures the luminosity difference between the BCG and the second brightest galaxy, we estimate it using the r-band magnitudes of BCG1 and BCG2.

To estimate the parameters (excluding $\Delta m_{12}$ and the offset indicators), we need to define the position and size of the region for assessing cluster shape. These regions are selected to optimise cluster segregation as disturbed or relaxed, with thresholds used for classification. While X-ray studies have explored cluster morphology using various centre definitions and apertures, SZ studies, particularly with high-resolution like NIKA2, are less common. To accurately classify the morphological indicators, we compare our results with simulation-based analyses. In this work, we adopt the same strategy used in \citet{c18}, using the SZ (or X-ray) centroids as the centre for the analysis. For the geometry and size of the apertures, when required, we use circular apertures with radii expressed as a fraction of $R_{500}$ (adopted from \citetalias{hass}: to have a standard value of $R_{500}$ compatible with the various estimates from this work, reported in Section \ref{masseval}), for further details, see \cite{luca}. In terms of the actual images used to estimate the morphological parameters: we use the PS-masked, $150 \, \rm GHz$ NIKA2 map, smoothed with a $10^{\prime \prime}$ Gaussian kernel (top-right panel of Fig. \ref{actnk}) for the SZ, and the background-subtracted and PS-in-painted photon counts map for the X-ray. To account for statistical uncertainties in the morphological parameters due to the noise in the map we use Monte Carlo realisations of the SZ and and X-ray maps. For the SZ, we simulate 1000 maps where the cluster signal is represented by the spherical model corresponding to the best-fit model obtained in Section \ref{pressureprof} and the noise is realised each time using the observed noise power spectrum in the ACTJ0240 NIKA2 $150 \, {\rm GHz}$ map. To simulate the effect of the shot noise in the indicators in X-ray, we estimate the expected photon counts from the cluster using the wavelet map. Then the count map is resampled, assuming a Poissonian distribution with these expected counts for each pixel. The morphological metrics are re-estimated for these maps and we consider the difference between the 84th percentile and the median (the median and the 16th percentile) of the parameter distributions as the positive (negative) error on the morphological parameters. In Table \ref{actmorpho}, we summarise our estimates for the morphological parameters in all three wavelength bands.

\begin{table*}
\centering
\begin{tabular}{|cccccc|}
\hline

 Image band & Parameter & Parameter description & Aperture & Value & $P_{\rm dis} $  \\
 &  &  &  [$R_{500}$] &  & [\%]  \\

\hline
Millimeter & $A$ & Asymmetry & 1.00 & $0.78\pm 0.03$ & $98.9\substack{+1.1\\-26.4}$  \\
 
 (SZ)  & $S$ & Strip & 1.00 & $0.29\substack{+0.20\\-0.15}$ & $98.6\substack{+1.4\\-27.7}$ \\

& $w$ & Centroid shift & 0.75 & $0.01\substack{+0.005\\-0.004}$ & $92.1\substack{+4.6\\-15.0}$ \\

& $\log (P_3/P_0)$ & Power ratio & 0.50 & $-5.98\substack{+0.14\\-0.16}$ & $99.2\substack{+0.6\\-0.9}$ \\

& $c$ & Light concentration ratio & 0.05-0.25 & $0.04\pm 0.002$ & $98.5\substack{+0.5\\-0.8}$  \\
 
& $M_{SZ}$ & Combined & -- & $1.55\substack{+0.13\\-0.12}$ & --   \\

& $\Delta_{BCG1-y_c}/R_{500}$ & BCG1-Centroid offset & -- & 0.33 & -- \\

& $\Delta_{BCG2-y_c}/R_{500}$ & BCG2-Centroid offset & -- & 0.19 & --  \\
\hline

X-ray & $A$ & Asymmetry & 0.75 & $0.75\pm 0.02$ & $91.8\substack{+0.7\\-1.0}$ \\

& $S$ & Strip & 1.00 & $0.60\substack{+0.06\\-0.04}$ & $98.5\substack{+0.5\\-1.4}$\\

& $w$ & Centroid shift & $0.75$ & $0.0049\substack{+0.0035\\-0.0010 }$ & $52.0\substack{+15.7	\\-38.8}$  \\

& $\log (P_3/P_0)$ & Power ratio & 0.25 & $-5.59\substack{+0.51\\-0.47}$ & $95.5\substack{+3.3\\-6.1}$  \\

& $c$ & Light concentration ratio & 0.025-0.25 & $0.0066\substack{+0.0023\\-0.0024}$ & $99.8\substack{+0.2\\-0.5}$ \\
 
& $M_X$ & Combined & -- & $0.94\substack{+0.24\\-0.13}$ & --   \\

& $\Delta_{BCG1-X_c}/R_{500}$ & BCG1-Centroid offset & -- & 0.27 & -- \\

& $\Delta_{BCG2-X_c}/R_{500}$ & BCG2-Centroid offset & -- & 0.24 & --  \\
\hline

Optical & $\Delta m_{12}$ & Magnitude gap (r band) & -- & 0.038 & -- \\

\hline

\end{tabular}
\caption{Morphological parameters inferred for ACTJ0240 at 3 different bands with the corresponding probability, $P_{\rm dis}$, that the inferred value is associated with a cluster that is not relaxed.}
\label{actmorpho}
\end{table*}

To determine if the cluster is relaxed, we compare our results with the parameter distributions from \citet{luca}. We then calculate the probability, $P_{\rm rel}$, of observing a simulated relaxed cluster in The300 with a morphological indicator as extreme/higher (or lower for $c$). Then our estimates $x^{obs}_i$:
\begin{equation}
    P = P_{\rm rel} (x_{\rm The300} \ge x_{obs}) = N_{rel}(x_{\rm The300} \ge x_{obs})/N_{rel}^{tot},
    \label{eq:prob}
\end{equation}
where $N_{rel}$ ($N_{rel}^{tot}$) is the (total) number of simulated relaxed clusters for the indicator $x$ (and the inequalities should be reversed in Eq. \ref{eq:prob} for the $c$ indicator). Thus, the probability for the cluster to not be relaxed is $P_{\rm dis} = 1-P_{\rm rel}$. We provide these probabilities for all the relevant morphological indicators, also in  Table \ref{actmorpho}. The errors on the probabilities are the propagated errors from the morphological parameters themselves.

Based on The300 results, all SZ and X-ray indicators, except the X-ray centroid shift, suggest a disturbed cluster with $P_{\rm dis}$ above 90\%. The combined estimates, $M_{SZ} \sim 1.55$ and $M_X \sim 0.94$, are well above the zero threshold for a relaxed cluster. The offsets between the X-ray ($\Delta_{BCG-X_c}/R_{500} > 0.24$) and SZ ($\Delta_{BCG-y_c}/R_{500} > 0.19$) centroids and the two BCGs also exceed the typical thresholds \citep[$\Delta_{BCG-c}/R_{500}\sim 0.05-0.07$ in][]{luca}. Although BCGs may not always align with the halo density peak, especially in disturbed clusters, where they could be displaced away during mergers or be related to satellite halos \citep[][and reference therein]{Hoshino2015, Cui2016, DePropris2021, Seppi2023}, the magnitude gap ($\Delta m_{12} = 0.038$) further confirms the cluster being in a disturbed state, as it is well below the $\Delta m_{12}>1.0$ threshold for relaxed clusters \citep{lopes}. Overall, morphological indicators across all bands quantitatively confirm that the cluster is not in a relaxed state.

\subsection{Pressure profiles}

The ICM pressure profiles are influenced by the cluster dynamical state. For instance, in \citetalias{upp}, the authors used the {\sc REXCESS} cluster sample \citep{upp_cc} to derive an average dimensionless universal pressure profile (UPP) and separate dimensionless profiles for cool-core (CC) and morphologically disturbed (MD) clusters, which are varied from each other. To make a similar comparison, we apply the best-fit parameters from their work to describe the physical pressure profile $P(r)$ of ACTJ0240 corresponding to the dimensionless profiles in \citetalias{upp}, using Eq. 13 of their work. For this estimation, we use $M_{500} = (3.3 \pm 0.8) \times 10^{14} \ h_{70}^{-1}\mathrm{M}_{\odot}$ and $z=0.62$ from \citetalias{hass}. While comparing these profiles to the ones we estimate we need to recall that sample in \citetalias{upp} is below the redshift of 0.2 and selected in the X-ray, whose selection biases might be different than that of an SZ-selected, and hence, the LPSZ sample. To draw a fair comparison in that aspect, we also adopt the best-fit values of the average pressure profiles derived using SZ samples. We utilise the results from \cite{pact2}, where the profile is derived using 31 clusters detected in both the Planck and ACT surveys in the redshift range between 0.16 and 0.70 (hereafter called PACT), and \cite{melin}, which uses 461 clusters detected jointly by the South Pole Telescope (SPT) and Planck (hereafter referred to as PSPT).

\begin{figure}
   \centering
   \includegraphics[width=\columnwidth]{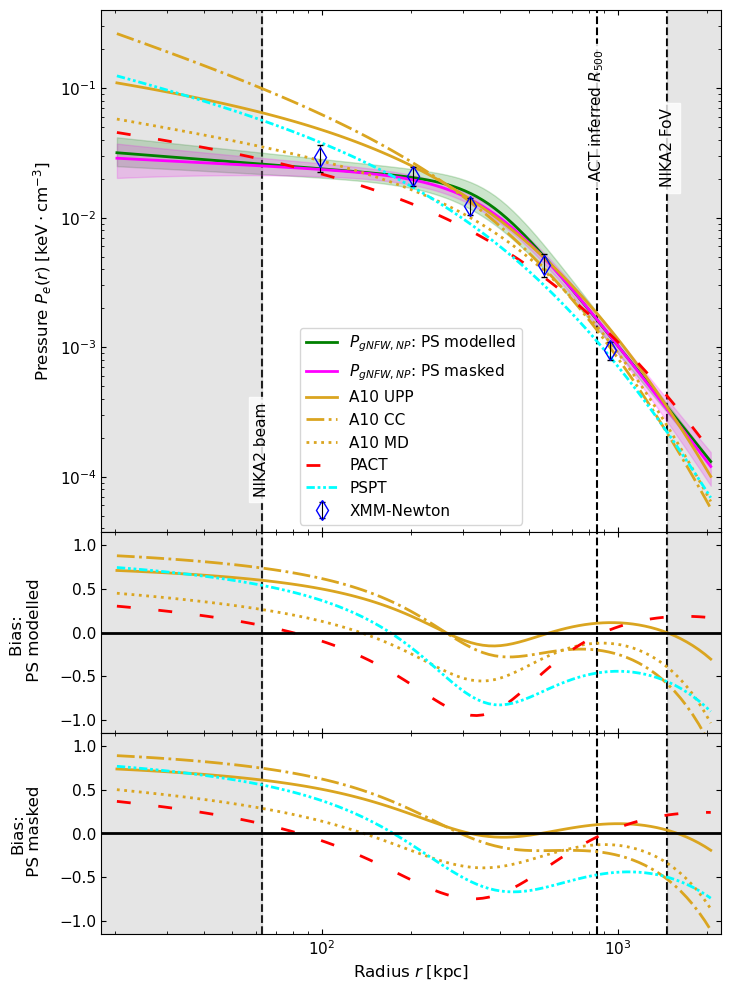}

   \caption{Comparison of NIKA2 SZ and \textit{XMM-Newton} inferred pressure profiles estimated in this work with various universal pressure profile models. The solid (green and magenta) lines, blue empty markers, and vertical lines are the same as Fig. \ref{pressureprof}. The solid, dotted-dashed, and dotted mustard lines represent a physical pressure profile representing ACTJ0240, corresponding to the average dimensionless UPP, profile for the CC sub-sample, and profile for the MD sub-sample, respectively, in \citetalias{upp}. The dashed red line and dashed-dotted cyan line correspond to the average pressure profile models PACT (see \citealt{pact2}) and PSPT (see \citealt{melin}), respectively. The shaded grey areas depict the regions where we have extrapolated our $P_{\rm gNFW,NP}$ best-fit models from the non-shaded area. The \textit{middle panel} (\textit{bottom panel}) shows the bias of our $P_{\rm gNFW,NP}$ profiles with respect to the various pressure profile models.}
    \label{uppcomp}%
\end{figure}

The resulting physical pressure profiles, equivalent to ACTJ0240 hypothetically belonging to a population representing the UPP, the CC sub-ample, and the MD sub-sample, PACT and PSPT average pressure profiles are shown in Fig. \ref{uppcomp} with solid, dotted-dashed, and dotted mustard lines, red dashed line, and cyan dashed-dotted line, respectively. The middle and bottom panels show the bias of our profiles with respect to the five best-fit models from the literature, where we define the bias as $(Model-P_{\mathrm{gNFW,NP}})/Model$. It is evident that the SZ pressure profiles $P_{\mathrm{gNFW,NP}}$ (solid green and magenta curve) are not well described by the UPP, nor the CC and MD pressure profile. On the other hand, the X-ray pressure points are completely consistent, within their own $1 \sigma$ limits, with a disturbed pressure profile for the cluster. Our pressure profiles are also not well described by the average pressure profiles PACT and PSPT, which are closer to us in terms of the observable used and the sample redshift range. Some discrepancies may stem from our fitting of all five gNFW parameters, whereas other studies fixed at least one, which, depending on its correlation with other free parameters, can significantly change the best-fit profile shape. Additionally, we only show the bias relative to the best-fit universal profiles, accounting for the dispersion in these models could, and most likely does, alleviate our discrepancies. Nonetheless, no universal profile matches our results, with ACTJ0240 showing a flattened pressure profile in the core, most similar to the MD model, suggesting it is likely a disturbed cluster.

\subsection{Impact on HSE mass estimate}

HSE masses are typically biased low compared to true cluster masses (\citealt{allen}). However, the disturbed state of ACTJ0240 likely increases this bias (\citealt{giulia}). To account for this additional bias, or the higher `scatter' in HSE mass due to the cluster being disturbed, we use bias value estimates from \cite{giulialatest}. That work indicates that while the median HSE bias is around $10\%$ for both relaxed and disturbed clusters, the scatter is greater for disturbed ones, with bias ranging from $20\%$ to $-10\%$ compared to $18\%$ to $0\%$ for relaxed clusters. Consequently, we re-estimate the associated systematic errors on cluster mass ($M_{500}$) to be $4.15\substack{+0.77\\-0.38} \times 10^{14} \, \mathrm{M}{\odot}$ when modelling the point source and $4.25\substack{+0.85\\-0.43} \times 10^{14} \, \mathrm{M}{\odot}$ when masking it. Unlike Section \ref{masseval}, where errors include both statistical and systematic uncertainties (e.g., PS contamination), these errors are purely systematic, reflecting the dispersion in mass estimates due to the higher HSE bias in disturbed clusters. Given the two errors have a similar order, this dynamical state-induced error must be included in the HSE mass bias error, when calibrating the LPSZ scaling law.

\section{Discussions and conclusions}
\label{actdiscussion}

In this work, we have produced maps and radial ICM profiles for galaxy cluster ACTJ0240 using multi-wavelength data, with a special focus on the novel, high-resolution SZ data from the NIKA2 LPSZ survey, and have estimated, for the first time, several morphological indicators of the cluster using high-resolution NIKA2 LPSZ image. We summarise and discuss our takeaways as follows:

\begin{itemize}
    \item \textbf{Pressure profiles:} While estimating the pressure profiles for ACTJ0240, we faced a unique challenge due to the lack of external information on the contaminating point source (PS). We addressed this using three different approaches that we defined earlier. In the NP model, all three approaches yielded consistent results within $1\sigma$, though the first case, where we are blind to the PS, understandably underestimated the pressure in the affected radial bin. Due to this dip, we could not constrain the gNFW model to fit on the NP data. However, the gNFW fits in the other two cases were consistent with each other and with the NP profiles, accurately fitting the cluster ICM up to $r\lesssim 800 \, {\rm kpc}$. Beyond this, the NP profiles fell more steeply than their gNFW fits due to factors essential for applying HSE to a disturbed cluster. Comparing these pressure profiles with the \textit{XMM-Newton} data showed consistency with the gNFW fits on the NP data, highlighting the complementarity of SZ and X-ray observations in mapping the cluster ICM.

    \item \textbf{HSE mass:} We estimated under the HSE assumption, the mass profile of the cluster for two scenarios: masking the contaminating PS and modelling it. The HSE mass profiles are consistent with each other and the X-ray only results, within $1\sigma$ uncertainties. The estimated mass values are also consistent with the estimates from previous studies mentioned in Section \ref{masseval}, with NIKA2 offering improved precision on mass, when we mask the PS.

    \item \textbf{Cluster dynamical state and its impact on mass estimates:} Qualitatively, we created a multi-band image incorporating SZ, X-ray, and optical data. The images reveal a disturbed cluster with misaligned centres and non-overlapping gas and galaxy distributions. SZ tail structures correspond to cold gas clumps in the X-ray and higher galaxy density in the optical, suggesting in-fall of matter from the southwest. Quantitatively, we estimated various morphological parameters (Table \ref{actmorpho}) and determined the probability for the cluster to be disturbed as $P_{\rm dis}>90\%$. The disturbed state leads to an added systematic error on the HSE mass which is important for consideration during the calibration of the LPSZ $Y-M$ scaling law.
    
\end{itemize}

In conclusion, we have created a multi-wavelength view of cluster ACTJ0240, with a prime focus on the high-resolution SZ data from NIKA2. The analysis of the morphology on the multi-band images, along with its pressure profile, indicates that the cluster is disturbed. This is the first case of analysing the cluster dynamical states using the NIKA2 LPSZ data of the cluster. This study emphasises the ability of the LPSZ to analyse and deal with a unique, disturbed cluster system while improving the statistical precision of mass estimation by the inclusion of resolved NIKA2 data.

\begin{acknowledgements} We would like to thank the IRAM staff for their support during the campaigns. The NIKA2 dilution cryostat has been designed and built at the Institut Néel. In particular, we acknowledge the crucial contribution of the Cryogenics Group, and in particular Gregory Garde, Henri Rodenas, Jean-Paul Leggeri, Philippe Camus. This work has been partially funded by the Foundation Nanoscience Grenoble and the LabEx FOCUS ANR-11-LABX-0013. This work is supported by the French National Research Agency under the contracts ``MKIDS'', ``NIKA'' and ANR-15-CE31-0017 and in the framework of the ``Investissements d'avenir'' program (ANR-15-IDEX-02). This work has been supported by the GIS KIDs. This work has benefited from the support of the European Research Council Advanced Grant ORISTARS under the European Union’s Seventh Framework Programme (Grant Agreement no. 291294). A.P., M.D.P., A.F. and R.W. acknowledge financial support from PRIN 2022 (Mass and selection biases of galaxy clusters: a multi-probe approach - n. 20228B938N) and Sapienza Università di Roma, thanks to Progetti di Ricerca Medi 2022, RM1221816758ED4E. E.P. and M.-M.E. acknowledge support from the French Agence Nationale de la Recherche (ANR), under grant ANR-22-CE31-0010. FDL acknowledges financial contribution from PrinMUR 2022, supported by Next Generation EU (n.20227RNLY3 The concordance cosmological model: stress-tests with galaxy clusters), Fondazione ICSC, Spoke 3: Astrophysics and Cosmos Observations, National Recovery and Resilience Plan (Piano Nazionale di Ripresa e Resilienza, PNRR) Project ID CN00000013 `Italian Research Center on High Performance Computing, Big Data and Quantum Computing' funded by MUR Missione 4 Componente 2 Investimento 1.4: Potenziamento strutture di ricerca e creazione di ``campioni nazionali di R\&S (M4C2-19)" - Next Generation EU (NGEU), and support from INFN through the InDark initiative. R.A. acknowledges support from the Programme National Cosmology et Galaxies (PNCG) of CNRS/INSU with INP and IN2P3, co-funded by CEA and CNES. R.A. was supported by the French government through the France 2030 investment plan managed by the National Research Agency (ANR), as part of the Initiative of Excellence of Université Côte d'Azur under reference number ANR-15-IDEX-01. A. R. acknowledges financial support from the Italian Ministry of University and Research - Project Proposal CIR01\_00010.
\end{acknowledgements}
%
%
\bibliographystyle{aa} 
\bibliography{references}

\end{document}